\documentclass[fleqn,usenatbib,onecolumn]{mnras}
\usepackage{newtxtext,newtxmath}

\usepackage[T1]{fontenc}
\usepackage{ae,aecompl}

\usepackage{epsfig}
\usepackage{graphicx}
\usepackage{amsmath}
\usepackage{amssymb}
\usepackage{mathrsfs}
\usepackage{mathtools}
\usepackage{hyperref}

\title[Instability of warped discs]{Instability of warped discs}
\author[Do\u{g}an, Nixon et al.]{
S.~Do\u{g}an$^{1,2}$\thanks{suzan.dogan@ege.edu.tr}, C.~J.~Nixon$^{2}$\thanks{cjn@leicester.ac.uk }, A.~R.~King$^{2,3,4}$ \& J.~E.~Pringle$^{2,5}$ \vspace{0.1in}\\
$^{1}$University of Ege, Department of Astronomy \& Space Sciences, Bornova, 35100, ${\dot {\rm I}}$zmir, Turkey\\
$^{2}$Theoretical Astrophysics Group, Department of Physics and Astronomy, University of Leicester, Leicester, LE1 7RH, UK\\
$^{3}$Anton Pannekoek Institute, University of Amsterdam, Science Park 904, 1098 XH Amsterdam, Netherlands\\
$^{4}$Leiden Observatory, Leiden University, Niels Bohrweg 2, NL-2333 CA Leiden, Netherlands\\
$^{5}$Institute of Astronomy, Madingley Road, Cambridge, CB3 0HA, UK
}

\date{Draft version, \today.}
\pubyear{2017}
\pagerange{\pageref{firstpage}--\pageref{lastpage}}
\begin{document}
\label{firstpage}
\maketitle

\begin{abstract}
Accretion discs are generally warped. If a warp in a disc is too large, the disc can `break' apart into two or more distinct planes, with only tenuous connections between them. Further if an initially planar disc is subject to a strong differential precession, then it can be torn apart into discrete annuli that precess effectively independently. In previous investigations, torque-balance formulae have been used to predict where and when the disc breaks into distinct parts. In this work, focusing on discs with Keplerian rotation and where the shearing motions driving the radial communication of the warp are damped locally by turbulence (the `diffusive' regime), we investigate the stability of warped discs to determine the precise criterion for an isolated warped disc to break. We find and solve the dispersion relation, which in general yields three roots. We provide a comprehensive analysis of this viscous-warp instability and the emergent growth rates and their dependence on disc parameters. The physics of the instability can be understood as a combination of (1) a term which would generally encapsulate the classical Lightman-Eardley instability in planar discs (given by $\partial(\nu\Sigma)/\partial\Sigma < 0$) but is here modified by the warp to include $\partial(\nu_1|\psi|)/\partial|\psi| < 0$ and (2) a similar condition acting on the diffusion of the warp amplitude given in simplified form by $\partial(\nu_2|\psi|)/\partial|\psi| < 0$. We discuss our findings in the context of discs with an imposed precession, and comment on the implications for different astrophysical systems.
\end{abstract}


\section{Introduction}
Accretion discs occur throughout astrophysics, from the birthsites of stars and planets to the feeding of supermassive black holes (SMBH) in galaxy centres. Through several different effects, each important in different contexts, these discs can warp. Discs can be formed in a warped configuration through e.g. chaotic infall on to protostellar discs \citep{Bate:2010aa} or supermassive black holes \citep{Lucas:2013aa}. Alternatively discs can warp if they orbit in a non-spherical gravitational potential which induces radial differential precession, e.g. the Lense-Thirring effect from a spinning black hole \citep{Lense:1918aa} or tides from a companion star \citep{Papaloizou:1995ab}. Further, there are warping instabilities which act on planar discs to drive a disc warp from an initially axisymmetric configuration with only a small perturbation, e.g. radiation warping \citep{Pringle:1996aa,Pringle:1997aa}, winds \citep{Schandl:1994aa} or resonant tidal effects \citep{Lubow:1992aa,Lubow:2000aa}.

Once a disc is warped, the local plane of the disc changes with radius. This creates a displacement of the high-pressure midplane region from one ring of the disc to its neighbour which oscillates with azimuth. At the descending (azimuthal angle $\phi=0$) and ascending ($\phi=\pi$) nodes, where the neighbouring rings cross, there is no displacement, whereas it is maximal at $\phi=\pi/2$ and (with opposite direction) at $3\pi/2$ \citep[see e.g. Fig. 1 of][]{Nixon:2016aa}. Thus, a fluid element orbiting in the disc feels a radial pressure gradient which oscillates on the local orbital frequency. The fluid element then exhibits epicyclic motion. In a near-Keplerian disc, the epicyclic frequency is close to the orbital frequency, creating a resonance between the motion and the forcing. The epicyclic motion communicates the disc warp radially, and if undamped will lead to the launch of a bending wave. However, in the presence of turbulence the wave will be damped and, if damped strongly enough such that the motions are dissipated locally, the behaviour takes on a diffusive character. Detailed discussions of the physics of these discs can be found in e.g. \cite{Papaloizou:1983aa,Papaloizou:1995aa,Ogilvie:1999aa,Ogilvie:2000aa,Lubow:2000aa,Lodato:2007aa} with some of the key points summarised in the review by \cite{Nixon:2016aa}.

To make progress, it is often useful to parameterise the turbulence in the disc \citep{Shakura:1973aa}. In planar, circular, thin discs this is usually done by taking the $R$-$\phi$ component of the vertically integrated stress tensor proportional to the vertically integrated pressure, with the constant of proportionality equal to a dimensionless parameter $\alpha$. Equivalently the kinematic viscosity $\nu = \alpha c_{\rm s} H$. For a warped disc, it is also necessary to determine the contributions to the radial communication of angular momentum from the $R$-$z$ and $\phi$-$z$ components of the stress tensor. This was first done using the linearised fluid equations by \cite{Papaloizou:1983aa}, and subsequently for the nonlinear case by \cite{Ogilvie:1999aa,Ogilvie:2000aa,Ogilvie:2013aa}. In these investigations, the damping of shearing motions is assumed to be isotropic, i.e. the same $\alpha$ causing dissipation from azimuthal shear, acts to dissipate vertical shear. As discussed in the above paragraph, damping of these motions inhibits radial communication of the misaligned component of angular momentum, leading to an inverse dependence between the $\alpha$ parameter and the strength of the torque. A simple, but informative, analytical description of this relation is provided in Section 4.1 of \cite{Lodato:2007aa}.

The isotropy of the $\alpha$ parameter used in almost all warped disc modelling has been the subject of much speculation within several recent MHD simulation studies that include MRI turbulence directly. These investigations reported a perceived departure from the behaviour of an $\alpha$ model \citep{Sorathia:2013aa,Sorathia:2013ab,Zhuravlev:2014aa,Morales-Teixeira:2014aa,Krolik:2015aa}. Unfortunately, none of these works compared their results directly with an $\alpha$ model. But there has been some recent progress on this issue. The $\alpha$ formalism provides a simple method for encapsulating the dominant effects of turbulence within the disc. For the general case of weak magnetic fields, the turbulence is found to be a local fluid property, with little power on scales larger than the disc thickness \citep[$\lesssim 10\%$;][]{Simon:2012aa}. As the disc cannot be warped on scales smaller than the disc thickness it seems reasonable to assume that the turbulence will not affect any particularly directed shear differently - suggesting the assumption of isotropy is reasonable. However, this cannot be taken as a given. A direct comparison between different codes is difficult due to the range of numerical techniques and different physics employed\footnote{To date only the SPH code {\sc phantom} \citep{Price:2017aa} has been tested for correctly modelling warped discs against the nonlinear theory of \cite{Ogilvie:1999aa} finding excellent agreement in the simulated parameter space.}. In spite of this, \cite{Nealon:2016aa} endeavored to match the parameters of the MHD simulation presented in \cite{Krolik:2015aa} and perform this calculation with an $\alpha$ viscosity playing the role of the disc turbulence to discern any differences found. However, no differences were found. Therefore \cite{Nealon:2016aa} directly confirms that, at least for the parameters simulated, the $\alpha$ model provides a suitable description of MRI turbulence for modeling warped discs. It is clear though, that the dynamics of $\alpha$ discs will diverge from MHD simulations if the magnetic field strength approaches dynamical importance and can provide local and large scale stresses which are comparable to the resonant Reynolds stress created by the warp itself. This is highly likely in e.g. simulations of MAD discs \cite[e.g.][]{McKinney:2013aa}, or in discs around highly magnetised stars, with the inner regions of discs in polars (AM Herculis) as an extreme example.

In this study we adopt the $\alpha$ formalism, in which recent numerical investigations show that accretion discs may break or tear into distinct planes when the viscosity is not strong enough to communicate the precession induced in the disc. Disc tearing has been shown to occur in discs inclined to the spin of a central black hole \citep{Nixon:2012ad,Nealon:2015aa}, in circumbinary discs around misaligned central binary systems \citep{Nixon:2013ab,Facchini:2013aa,Aly:2015aa} and in circumprimary discs misaligned with respect to the binary orbital plane \citep{Dogan:2015aa}. In these initial papers the criterion for disc tearing has been derived simplistically by comparing the viscous torque with the precession torque induced in the disc, or in the case of wavelike warp propagation the wave travel time was compared to the precesion timescale. Initially, the precession torque was compared to the torque arising from azimuthal shear \citep{Nixon:2012ad}. For small $\alpha$ and small disc inclination angles, the simulations of \cite{Dogan:2015aa} showed that the inclusion of the effective viscosity arising from vertical shear was required. However, as this term depends on the disc structure through the warp amplitude $\left|\psi\right|$, which cannot be determined {\it a priori}, it is not very useful as a predictive criterion.

Further to these simulations of disc tearing, it is also possible for warped discs to break without any external forcing \citep{Lodato:2010aa}. For these isolated warped discs the evolution must be driven by the dependence of the effective viscosities on the disc structure. This instability occurs in a simpler environment than disc tearing, and likely underpins that behaviour. In this work we seek to understand the physics causing an isolated warped disc to break. We do this by performing a stability analysis of the warped disc equations, finding a dispersion relation and the instability criterion. As we shall see below, instability appears in the form of anti-diffusion which leads to a discontinuity in the disc angular momentum. This can be achieved by a discontinuity in the surface density or in the disc inclination. In the case of a flat disc, this is the familiar \citep{Lightman:1974aa} viscous disc instability, which underlies the thermal-viscous instability commonly thought to be the basis of dwarf nova outbursts \citep{Meyer:1982aa}, and when combined with central X-ray irradiation, of soft X-ray transient outbursts \citep{King:1998aa}. The thermal-viscous stability of a warped disc was briefly analysed by \cite{Ogilvie:2000aa}, who concluded that a disc that is Lightman-Eardley stable (like those we consider here) is usually viscously stable when warped except for $\alpha \lesssim 0.05$ and the warp exceeds a critical amplitude. For a warped disc anti-diffusion can lead to a separating of the disc into two or several discrete planes, which are connected by a series of sharply changing, low surface-density orbits.

The layout of the paper is as follows. In Section~\ref{sec:equations} we describe the governing equations and their assumptions. In Section~\ref{sec:analysis} we obtain the dispersion relation and derive the instability condition to determine a general criterion for discs to break. Further we provide numerical evaluation of the growth rates of the instability. In Section~\ref{sec:heuristic} we provide a simple physical picture of the instability. In Section~\ref{sec:tearing} we discuss our findings in the context of disc tearing where the disc is subject to an external torque. Finally in Section~\ref{sec:conclusions} we provide our conclusions.

\section{Warped disc evolution equations}
\label{sec:equations}

In this work we are considering an isolated, Keplerian, warped disc with $\alpha > H/R$. In the literature, there exist various notations for the underlying equations with some subtleties which we outline here. 

\subsection{Definitions from previous works}
Using the notation of \cite{Pringle:1992aa} the evolution equation for the disc angular momentum is
\begin{eqnarray}
\label{eq:dLdt}
 \frac{\partial \mathbfit{L}}{\partial t} & = & \frac{1}{R}
 \frac{\partial }{\partial R} \left\{ \frac{\left(\partial / \partial
   R \right) \left[\nu_{1}\Sigma R^{3}\left(-\Omega^{'} \right)
     \right] }{\Sigma \left( \partial / \partial R \right) \left(R^{2}
   \Omega \right)} \mathbfit{L}\right\} \\ \nonumber & &
 +~~{}\frac{1}{R}\frac{\partial}{\partial R}\left[\frac{1}{2} \nu_{2}R
   \left| \mathbfit{L} \right|\frac{\partial \mathbfit{l}}{\partial
     R} \right] \\ \nonumber & &
 +~~{}\frac{1}{R}\frac{\partial}{\partial R}
 \left\{\left[\frac{\frac{1}{2}\nu_{2}R^{3}\Omega \left|\partial
     \mathbfit{l} / \partial R \right| ^{2}}{\left( \partial /
     \partial R \right) \left( R^{2} \Omega \right)} +
   \nu_{1}\left(\frac{R \Omega^{'}}{\Omega} \right) \right]
 \mathbfit{L}\right\} \\ \nonumber & &
 +~~{}\frac{1}{R}\frac{\partial}{\partial R} \left(\nu_{3} R
 \left|\mathbfit{L} \right| \mathbfit{l} \times \frac{\partial
   \mathbfit{l}}{\partial R} \right)
\end{eqnarray}
where $\nu_1$ and $\nu_2$ are the effective viscosities, $\nu_3$ is the coefficient of a dispersive wavelike torque, $\Omega\left(R\right)$ is the orbital angular velocity of each annulus of the disc, $\Sigma\left(R,t\right)$ is the disc surface density, $\mathbfit{l}\left(R,t\right)$ is the unit angular momentum vector pointing perpendicular to the local orbital plane, and $\Omega^\prime = \partial\Omega/\partial R$. Also $\mathbfit{L} = \Sigma R^2 \Omega \mathbfit{l}$ is the angular momentum per unit area.

This equation (\ref{eq:dLdt}) contains four distinct terms responsible for radial communication of orbital angular momentum ($\nu_1$ term), radial communication of the misaligned component of angular momentum ($\nu_2$ term), an advective term resulting from the first two torques, and a precessional torque between misaligned rings ($\nu_3$ term). This equation was derived by \cite{Pringle:1992aa} from conservation of mass and angular momentum, but excluding the $\nu_3$ term which is not required by conservation. The $\nu_3$ term arises as the imaginary part of the complex diffusion coefficient in the fluid analysis of \cite{Papaloizou:1983aa}.

The linearised analysis of \cite{Papaloizou:1983aa} provides the leading order corrections to the coefficients $\alpha_i$ of the effective viscosities ($\nu_1$ and $\nu_2$) and the dispersive torque ($\nu_3$) in the limit of small $\alpha$ and small warp amplitude. Developing this theory further, \cite{Ogilvie:1999aa} and subsequently \cite{Ogilvie:2000aa} and \cite{Ogilvie:2013aa}, provide a nonlinear derivation of this equation valid for general $\alpha$ and warp amplitude. On the whole, these works confirm the above equation, but include the important effect that the local fluid properties have on the torque coefficients.

\cite{Ogilvie:1999aa} presents (\ref{eq:dLdt}) as\footnote{Note that both $r$ here and $R$ in (\ref{eq:dLdt}) represent the local cylindrical radius of the annulus.}
\begin{equation}
  \label{eq:dLdtO99}
  \frac{\partial}{\partial t}\left(\Sigma r^2\Omega\mathbfit{l}\right)
  + \frac{1}{r}\frac{\partial}{\partial r}\left(\Sigma \bar{\upsilon}_r r^3\Omega\mathbfit{l}\right)
  = \frac{1}{r}\frac{\partial}{\partial r}\left(Q_1\mathcal{I}r^2\Omega^2\mathbfit{l}\right)
  + \frac{1}{r}\frac{\partial}{\partial r}\left(Q_2\mathcal{I}r^3\Omega^2\frac{\partial \mathbfit{l}}{\partial r}\right)
  + \frac{1}{r}\frac{\partial}{\partial r}\left(Q_3\mathcal{I}r^3\Omega^2\mathbfit{l}\times\frac{\partial \mathbfit{l}}{\partial r}\right)\,,
\end{equation}
where $\bar{\upsilon}_r$ is the mean radial velocity, $\mathcal{I}$ is the azimuthally averaged second vertical moment of the density, and $Q_i$ are the dimensionless torque coefficients. The coefficient $Q_1$ represents the usual viscous torque tending to spin the ring up or down. Coefficient $Q_2$ represents the torque diffusing the disc tilt. This torque tends to align the ring with its neighbours and flatten the disc. Coefficient $Q_3$ represents the precession torque causing the rings misaligned with its neighbours to precess. The $Q_3$ term is responsible for the dispersive wave-like propagation of the warp in the non-Keplerian inviscid case, but as we shall see below is relatively unimportant for the usual Keplerian diffusive case. The equivalence between (\ref{eq:dLdt}) and (\ref{eq:dLdtO99}) above can be seen through the equation for the radial velocity
\begin{equation}
  \bar{\upsilon}_r = \frac{\left(\partial/\partial r\right)\left(\nu_1\Sigma r^3\Omega^\prime\right)-\frac{1}{2}\nu_2\Sigma r^3 \Omega\left|\partial\mathbfit{l}/\partial r\right|^2}{r\Sigma\left(\partial/\partial r\right)\left(r^2\Omega\right)}\,,
\end{equation}
and the definitions of the effective viscosities
\begin{equation}
  \nu_1=\frac{(-Q_1)\mathcal{I}\Omega}{q\Sigma}
\end{equation}
\begin{equation}
  \nu_2=\frac{2Q_2\mathcal{I}\Omega}{\Sigma}
\end{equation}
\begin{equation}
  \nu_3=\frac{Q_3\mathcal{I}\Omega}{\Sigma}
\end{equation}
  where
  \begin{equation}
  q=-\frac{d \ln \Omega}{d \ln r}\,.
\end{equation}
  In \cite{Ogilvie:1999aa} the coefficients $Q_i$ are calculated for a polytropic disc. However, \cite{Ogilvie:1999aa} does not provide a complete analysis as the behaviour of $\mathcal{I}$ with warp amplitude $\left|\psi\right| = r|\partial \mathbfit{l}/\partial r|$ is not calculated there, and thus there is no simple equation relating $\mathcal{I}$ with $\Sigma$. Instead this equation takes the form $\mathcal{I} = f(\left|\psi\right|)\Sigma c_{\rm s}^2/\Omega^2$. This additional function $f(\left|\psi\right|)$ is missing from the calculations in \cite{Nixon:2012aa}, who assumed $\mathcal{I} = \Sigma c_{\rm s}^2/\Omega^2$.

\cite{Ogilvie:2000aa} provides a more complete thermal treatment including viscous heating and radiative cooling. This included  an azimuthally dependent vertical treatment of the warp and thus the effect of vertical squeezing of the disc by the warp. In \cite{Ogilvie:2000aa}, the evolution equation is presented as above (\ref{eq:dLdtO99}), complemented by a relation between $\mathcal{I}$ and $\Sigma$. In this relation, the dependence on the warp is included with a $Q_5$ parameter.

Finally, more recently, \cite{Ogilvie:2013aa} presented a further set of equations entirely consistent with those above, but with a subtle change to the definitions. They give the evolution equations as general conservation equations, with the internal torque defined separately. Putting them together yields
\begin{equation}
  \label{eq:dLdtOL13}
  \frac{\partial}{\partial t}\left(\Sigma r^2\Omega\mathbfit{l}\right)
  + \frac{1}{r}\frac{\partial}{\partial r}\left(\Sigma \bar{\upsilon}_r r^3\Omega\mathbfit{l}\right)
  = \frac{1}{r}\frac{\partial}{\partial r}\left(Q_1\Sigma c_{\rm s}^2 r^2\mathbfit{l}\right)
  + \frac{1}{r}\frac{\partial}{\partial r}\left(Q_2\Sigma c_{\rm s}^2 r^3\frac{\partial \mathbfit{l}}{\partial r}\right)
  + \frac{1}{r}\frac{\partial}{\partial r}\left(Q_3\Sigma c_{\rm s}^2 r^3\mathbfit{l}\times\frac{\partial \mathbfit{l}}{\partial r}\right)\,.
\end{equation}
Inspection of (\ref{eq:dLdtOL13}) and (\ref{eq:dLdtO99}) appears to show that $\mathcal{I} = \Sigma c_{\rm s}^2/\Omega^2$, and that the $\left|\psi\right|$ dependence of $\mathcal{I}$ has been ignored. However, this is not the case. Instead this dependence has been included through a change in definition of the $Q_i$ coefficients between these works. We have been unable to find direct reference to this change of definition. The motivation is clear though, and provided by \cite{Ogilvie:2013aa} who remark that this form of the torque is natural for the isothermal discs under consideration there (and here).

\subsection{Definitions in this work}
\label{sec:defshere}

In this work we adopt the following formalism based on the above discussion. This is in line with \cite{Ogilvie:2013aa} and is most relevant to the isothermal discs we consider here.

We note that while we are using the coefficients from \cite{Ogilvie:2013aa} which are calculated assuming a locally isothermal equation of state, our results are more widely applicable. For example, \cite{Ogilvie:2000aa}, who calculates the coefficients including an explicit energy equation, assumes that the disc is highly optically thick. This means that the azimuthal variations in physical variables e.g. $\rho$, $T$ are essentially adiabatic ($t_{\rm therm} \gg t_{\rm dyn}$). It follows that in this highly optically thick case, calculation of the coefficients assuming an isothermal equation of state and those employing a polytropic equation of state or including an explicit energy equation differ only in the adiabatic exponent $\gamma$. Thus we expect the physics in each case to differ only marginally. This appears to be confirmed by inspection of e.g. Figs. 3 \& 4 of \cite{Ogilvie:1999aa} and Figs. 2 \& 3 of \cite{Ogilvie:2000aa}. We conclude from this that the calculations we present below, which are exact for the isothermal case (and thus also relevant to the optically thin case), are also applicable to the highly optically thick case. If by chance $t_{\rm therm} \sim t_{\rm dyn}$ then the azimuthal variations are no longer adiabatic, but we expect the basic physics to remain unchanged.\footnote{In a steady disc, $t_{\rm therm}\sim \alpha^{-1}t_{\rm dyn}$ \citep{Pringle:1981aa}, and so in general $t_{\rm therm} \gtrsim t_{\rm dyn}$ with rough equality only for $\alpha \approx 1$.}

For our purposes, the conservation of mass equation is
\begin{equation}
 \label{eq:conservation1}
\frac{\partial \Sigma}{\partial t}+\frac{1}{r}\frac{\partial}{\partial r}(r\bar{\upsilon}\Sigma)=0\,,
\end{equation}
and the conservation of angular momentum equation is
\begin{equation}
\label{eq:conservation2}
  \frac{\partial }{\partial t}\left(\Sigma r^2\Omega \mathbfit{l}\right)
  +\frac{1}{r}\frac{\partial}{\partial r}(\Sigma \bar{\upsilon}_r r^3\Omega \mathbfit{l})
  =\frac{1}{r}\frac{\partial}{\partial r}\left(Q_1\Sigma c_{\rm s}^2 r^2\mathbfit{l}\right)
  + \frac{1}{r}\frac{\partial}{\partial r}\left(Q_2\Sigma c_{\rm s}^2 r^3\frac{\partial \mathbfit{l}}{\partial r}\right)
  + \frac{1}{r}\frac{\partial}{\partial r}\left(Q_3\Sigma c_{\rm s}^2 r^3\mathbfit{l}\times\frac{\partial \mathbfit{l}}{\partial r}\right)\,.
\end{equation}
The effective viscosities are related to the $Q_i$ coefficients by
\begin{equation}
  \nu_1 = (-Q_1)c_{\rm s}^2/q\Omega
\end{equation}
\begin{equation}
  \nu_2 = 2Q_2 c_{\rm s}^2/\Omega
\end{equation}
\begin{equation}
  \nu_3 = Q_3 c_{\rm s}^2/\Omega\,,
\end{equation}
and the $Q_i$ coefficients are evaluated using a code provided by Ogilvie following the calculations in \cite{Ogilvie:2013aa}.

By also defining the specific angular momentum
\begin{equation}
  h = r^2\Omega\,,
\end{equation}
with $h^\prime = {\rm d} h/{\rm d} r$, and substituting in the radial velocity, we have
\begin{equation}
  \label{eq:here}
  \frac{\partial }{\partial t}\left(\Sigma r^2\Omega \mathbfit{l}\right)
  =\frac{1}{r}\frac{\partial}{\partial r}\left[Q_1\Sigma c_{\rm s}^2 r^2\mathbfit{l} + Q_2\Sigma c_{\rm s}^2 r^3\frac{\partial \mathbfit{l}}{\partial r} + Q_3\Sigma c_{\rm s}^2 r^3\mathbfit{l}\times\frac{\partial \mathbfit{l}}{\partial r} - \left(\frac{\partial}{\partial r}\left[Q_1\Sigma c_{\rm s}^2 r^2\right] - Q_2\Sigma c_{\rm s}^2r\left|\psi\right|^2\right)\frac{h}{h^\prime}\mathbfit{l}\right]\,.
\end{equation}
This equation, supplemented by the $Q_i$ coefficients from \cite{Ogilvie:2013aa} is what we will work with for the remainder of this paper.

\section{Stability analysis of the warped disc equations}
\label{sec:analysis}

\subsection{Dispersion Relation}
We consider the local stability of an isolated warped disc, with $\alpha > H/R$. From (\ref{eq:here}) we define the internal torque components $g_i(r,\Sigma,\alpha,\alpha_{\rm b},q,|\psi|)$ that premultiply the dimensionless basis vectors $(\mathbfit{l},r\partial \mathbfit{l}/\partial r,r\mathbfit{l}\times\partial \mathbfit{l}/\partial r)$ as \citep{Ogilvie:2000aa}
\begin{equation}
\label{eq:gi}
g_i=Q_i(\alpha,\alpha_b,q,|\psi|)\Sigma c_s^2r^2.
\end{equation}
In this work we assume that the shear viscosity parameter $\alpha$ is a constant, and that the bulk viscosity parameter $\alpha_{\rm b} = 0$. We focus on Keplerian discs with $q=3/2$, and thus $Q_i = Q_i(\alpha,\left|\psi\right|)$. 

Using (\ref{eq:gi}), we can rewrite (\ref{eq:here}) as
\begin{equation}
 \label{eq:conservation4}
h\frac{\partial}{\partial t}(\Sigma \mathbfit{l})=\frac{1}{r}\frac{\partial}{\partial r}\left[\textit{g}_1\mathbfit{l}+\textit{g}_2r\frac{\partial \mathbfit{l}}{\partial r}+\textit{g}_3r\mathbfit{l}\times\frac{\partial \mathbfit{l}}{\partial r}-\left( \frac{\partial\textit{g}_1}{\partial r}-\frac{\textit{g}_2|\psi|^2}{r}\right)\frac{h}{h'}\mathbfit{l}\right].
\end{equation}
Following \cite{Ogilvie:2000aa}, we consider the stability of any solution of (\ref{eq:conservation4}) with respect to linear perturbations ($\delta\Sigma, \delta\mathbfit{l}$). The perturbed form of the angular momentum equation is
\begin{equation}
 \label{eq:conservation5}
\begin{array}{ll}
h\displaystyle\frac{\partial}{\partial t}(\delta\Sigma \mathbfit{l}+\Sigma \delta\mathbfit{l})=&{\displaystyle\frac{1}{r}\frac{\partial}{\partial r}\Bigg[\delta\textit{g}_1\mathbfit{l}+\textit{g}_1\delta\mathbfit{l}+\delta\textit{g}_2r\displaystyle\frac{\partial \mathbfit{l}}{\partial r}+\textit{g}_2r\displaystyle\frac{\partial \delta\mathbfit{l}}{\partial r}+\delta\textit{g}_3r\mathbfit{l}\times\frac{\partial \mathbfit{l}}{\partial r}+\textit{g}_3r\delta\mathbfit{l}\times\frac{\partial \mathbfit{l}}{\partial r}+\textit{g}_3r\mathbfit{l}\times\frac{\partial \delta\mathbfit{l}}{\partial r}}\\
\\
&{-\left( \displaystyle\frac{\partial\delta\textit{g}_1}{\partial r}-\displaystyle\frac{\delta(\textit{g}_2|\psi|^2)}{r}\right)\displaystyle\frac{h}{h'}\mathbfit{l}-\left( \displaystyle\frac{\partial\textit{g}_1}{\partial r}-\displaystyle\frac{\textit{g}_2|\psi|^2}{r}\right)\displaystyle\frac{h}{h'}\delta\mathbfit{l}\Bigg]}.
\end{array}
\end{equation}
Here the perturbed quantities are defined by
\begin{equation}
\delta\textit{g}_i=\frac{\partial\textit{g}_i}{\partial\Sigma}\delta\Sigma+\frac{\partial\textit{g}_i}{\partial\left|\psi\right|}\delta\left|\psi\right|
\end{equation}
and
\begin{equation}
\delta|\psi|=\frac{r^2}{|\psi|}\frac{\partial\mathbfit{l}}{\partial r}\cdot\frac{\partial\delta\mathbfit{l}}{\partial r}
\end{equation}
and $\mathbfit{l}\cdot\delta\mathbfit{l}=0$. We now search for solutions to the perturbations of the form
\begin{equation}
  \label{eq:perturbsol}
\exp\left(-i\int\omega dt + i\int k dr \right)
\end{equation}
where $\omega$ is the wave frequency and \emph{k} is the wavenumber. This requires $|kr|\gg1$, and while formally $k \rightarrow \infty$ is acceptable, the equations become invalid as $k$ approaches $1/H$ \citep{Ogilvie:2000aa}. Similarly $\left|\omega\right|$ must be large enough such that the perturbations grow or decay on timescales short compared to the background solution. We adopt an arbitrary scaling $|\delta \mathbfit{l}|=O(k^{-1})$, then $\delta \Sigma=O(1)$, $\delta |\psi|=O(1)$ and $\delta g_i=O(1)$, while $\omega=O(k^2)$ as expected for a diffusion equation \citep{Ogilvie:2000aa}. If we define an orthonormal basis ($\mathbfit{l}$, \textbf{\emph{m}}, \textbf{\emph{n}}) by $\mathbfit{l}\times\textbf{m}=\textbf{n}$ and $\delta \mathbfit{l}=\delta m\textbf{\emph{m}} + \delta n\textbf{\emph{n}}$ (where $r(\partial \mathbfit{l} / \partial r)=|\psi|\textbf{\emph{m}}$), we obtain the $\mathbfit{l}$, \textbf{\emph{m}} and \textbf{\emph{n}} components of (\ref{eq:conservation5}), at leading order in $k$, as follows:
\begin{equation}
-i\omega hr\delta \Sigma=k^2\frac{h}{h'}\delta\textit{g}_1
\end{equation}
\begin{equation}
-i\omega hr\Sigma\delta m=ik|\psi|\delta\textit{g}_2-k^2\textit{g}_2r\delta m+k^2\textit{g}_3r\delta n-ik|\psi|\frac{h}{rh'}\delta\textit{g}_1
\end{equation}
\begin{equation}
-i\omega hr\Sigma\delta n=ik|\psi|\delta\textit{g}_3-k^2\textit{g}_2r\delta n-k^2\textit{g}_3r\delta m
\end{equation}
\\
where $\delta|\psi|=ikr\delta m$. We can express $\delta g_i$ in terms of $\delta\Sigma$ and $\delta |\psi|$ by differentiating (\ref{eq:gi}), then write $\delta |\psi|$ in terms of $\delta m$. As a result, we get three linear equations with three unknowns ($\delta\Sigma$, $\delta m$, $\delta n$) which yield a coefficients determinant:
\begin{equation}
\label{eq:det}
\rm det
\renewcommand\arraystretch{1.5}
\begin{bmatrix}
    s-a Q_1      & -aQ_1' & 0  \\
    (Q_2-aQ_1) |\psi|       & s+Q_2+Q_2'|\psi|-aQ_1'|\psi| & - Q_3 \\
     Q_3|\psi|    & Q_3+Q_3'|\psi| & s+ Q_2
\end{bmatrix}
=0.
\end{equation}
Here, the prime on $Q_i$ represents differentiation with respect to $|\psi|$, $a = h/rh^\prime = {\rm d}\ln r/{\rm d}\ln h = 1/(2-q) = 2$ (for a Keplerian disc with $q=3/2$), $s$ is defined by
\begin{equation}
  s=-\frac{{\rm i}\omega}{\Omega}\Bigg(\frac{\Omega}{c_s k}\Bigg)^2\,,
\end{equation}
and we shall call $\Re[s]$ the dimensionless growth rate. $s$ is related to the physical growth rate ($\Re[-i\omega]$) through
\begin{equation}
  -{\rm i}\omega=s{\Omega}\Bigg(\frac{c_s k}{\Omega}\Bigg)^2.
\end{equation}
and the perturbations grow ($\Re[s] >0$) or decay ($\Re[s]<0$) as $\exp[\Re(-i\omega)t]$.

This shows that the physical growth rate is faster on smaller length scales as expected for a diffusion equation.  As the disc pressure scalelength is defined by $H_p\equiv c_s/\Omega$, the dimensionless growth rate can be rewritten as $s=-i(\omega/\Omega)(H_pk)^{-2}$. Therefore, $s\sim 1$ indicates strong instability on dynamical timescales and lengthscales $\sim H_{\rm p}$. We again note that validity of the equations requires $k \lesssim 1/H$. The determinant (\ref{eq:det}) gives a third order dispersion relation:
\\
\begin{equation}
\label{eq:dr}
\setlength\jot{12pt}
\begin{split}
    s^3&-s^2\left[aQ_1-2Q_2+|\psi|\Big(aQ'_1-Q'_2\Big)\right] \\
   & -s\left[2a{Q}_1{Q}_2-Q_2^2-Q_3^2+|\psi|\Big(a{Q}_1{Q}'_2-Q_2Q'_2-Q_3Q'_3\Big)\right]\\
   &-a\left[Q_1(Q_2^2+Q_3^2)+|\psi|\Big(Q_1Q_2Q'_2-Q_1'Q_2^2+Q_1Q_3Q'_3-Q_1'Q_3^2\Big)
   \right]=0.
\end{split}
\end{equation}
\\
The disc becomes unstable if any of the roots of (\ref{eq:dr}) has a positive real part, i.e. $\Re(s)>0$, as the perturbations then grow exponentially with time.

For completeness, we note, following the discussion of the thermodynamics in Section~\ref{sec:defshere} that the equivalent of (\ref{eq:dr}) for an analysis explicitly including the energy equation is given by \citep[cf.][]{Ogilvie:2000aa}
\begin{equation}
\label{eq:dr_alt}
\setlength\jot{12pt}
\begin{split}
    s^3&-s^2\left[apQ_1-2Q_2+|\psi|\Big(aQ'_1-Q'_2\Big)\right] \\
   & -s\left[2ap{Q}_1{Q}_2-Q_2^2-Q_3^2+|\psi|\Big(ap{Q}_1{Q}'_2-Q_2Q'_2-Q_3Q'_3 -a(p-1)Q'_1Q_2\Big)\right]\\
   &-ap\left[Q_1(Q_2^2+Q_3^2)+|\psi|\Big(Q_1Q_2Q'_2-Q_1'Q_2^2+Q_1Q_3Q'_3-Q_1'Q_3^2\Big)
   \right]=0.
\end{split}
\end{equation}
\\
where $p$ is the exponent of $\Sigma$ in $\nu\Sigma$ (or equivalently the exponent of $\Sigma$ in $\mathcal{I}$). In a planar disc, the usual thermal-viscous stability criterion \citep{Lightman:1974aa} is given by $p<0$. For the isothermal discs we consider here, we have $p=1$ (for which \ref{eq:dr} and \ref{eq:dr_alt} are identical). For Thomson scattering opacity, $p=5/3$, and for Kramers opacity, $p=10/7$, and as we have noted above the coefficients do not vary strongly between these cases. Therefore we expect the calculations we present here (with $p=1$ and coefficients derived from an isothermal equation of state) to hold at least qualitatively for a wide range of thermodynamic treatments.

\subsection{Instability Criterion and the Growth Rates}
\label{sec:criterion}

Before arriving at the full criterion (provided below) we first give the roots of (\ref{eq:dr}) for some simplified cases. We will see below that these cases are generally useful for both understanding the more complex cases, and as close approximations to the full solution.

\begin{figure}
  \begin{center}
        {\includegraphics[angle=270,scale=.38]{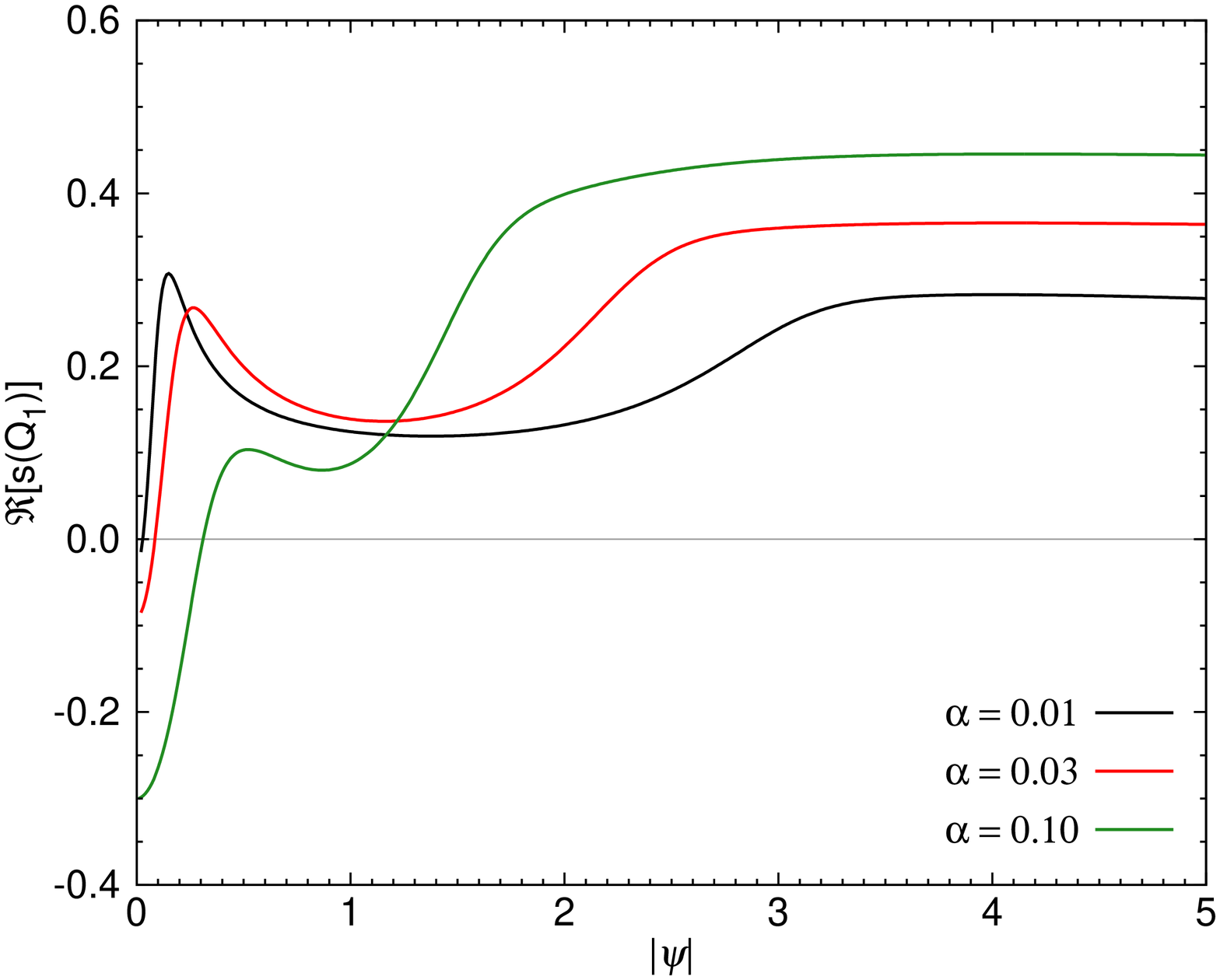}}
        {\includegraphics[angle=270,scale=.38]{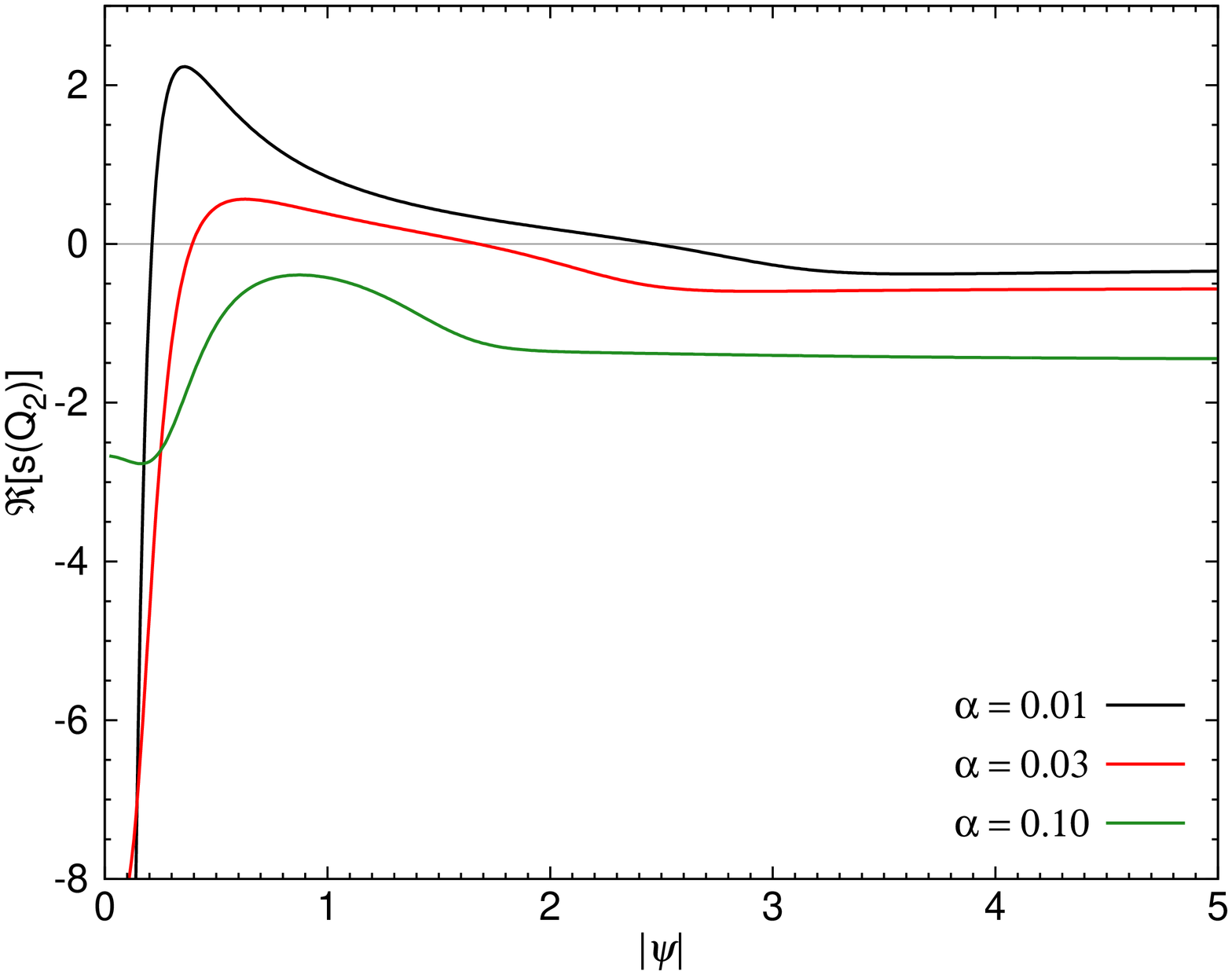}}
        \caption{Left hand panel: This shows the dimensionless growth rate $\Re[s(Q_1)]$ (equation \ref{eq:sq1}) as a function of dimensionless disc warp amplitude $\left|\psi\right|$ caused by the term $Q_1$ in the equations which governs radial angular momentum transfer, for three different values of the viscosity parameter $\alpha$. Instability occurs when $\Re[s] > 0$. We note that the maximal growth rates $\approx \Re[s]\Omega \sim 0.3\Omega$ are comparable with dynamical and can occur for small values of $\left|\psi\right|$, and remain unstable for large values of $\left|\psi\right|$. Right hand panel: This shows the dimensionless growth rate $\Re[s(Q_2)]$ caused by the action of the term $Q_2$ which governs the viscous flattening of the warped disc. We note that this term shows no instability for large values of $\alpha$. For smaller values of $\alpha$ there is a range of disc warp amplitude $\left|\psi\right|$ for which instability occurs. When it does occur, the timescale for instability can be short with growth rates $\approx \Re[s]\Omega \sim \Omega$ being dynamical. In both panels the grey line represents zero growth rate. We remind the reader that these illustrative calculations are based on the artificial assumption that two of the three $Q_{i}$ are set to zero.}
        \label{fig:simplified}
  \end{center}
\end{figure}

We label the roots as $s(Q_i)$, where the subscript $i$ denotes the torque coefficient(s) that we include in the solution. First, if we set all $Q_i$ terms in the dispersion relation to zero apart from $Q_1$ (i.e. $Q_2 = Q_3 = 0$), we find one root which is given as
\begin{equation}
\label{eq:sq1}
s(Q_1)= a\Big(Q_1 + Q'_1|\psi|\Big) = -a\frac{\partial}{\partial \left|\psi\right|}\left(-Q_1\left|\psi\right|\right)\,.
\end{equation}
For instability, one requires $\Re[s(Q_1)]>0$. For $\left|\psi\right|=0$, this corresponds to the usual criterion for viscous instability in a planar disc (given by $\partial(\nu\Sigma)/\partial\Sigma < 0$; \citealt{Lightman:1974aa}, although we note that the relevant term is simplified here as we have assumed a simple equation of state, cf equation 83 of \citealt{Ogilvie:2000aa}). However, in the warped case, the criterion is modified with an extra term determined by the behaviour of the coefficient $Q_1$ with warp amplitude.  Fig. \ref{fig:simplified} (left panel) shows the variation of the dimensionless growth rates of this mode, $\Re[s(Q_1)]$ with $|\psi|$ for $\alpha = 0.01, 0.03$ and $0.1$. The disc becomes unstable after some critical warp amplitude, $|\psi|_c$. The critical warp amplitude required for instability is very small for discs with low $\alpha$. The growth rates are a modest fraction $\gtrsim 0.1$ of dynamical. As a result, discs can easily become unstable for this case.

When we take into account only $Q_2$ (i.e. $Q_1 = Q_3 = 0$), then the solution of (\ref{eq:dr}) gives:
\begin{equation}
\label{eq:sq2}
s(Q_2)=-\Big(Q_2+Q'_2|\psi|\Big) = -\frac{\partial}{\partial \left|\psi\right|}\left(Q_2\left|\psi\right|\right)\,.
 \end{equation}
The disc becomes unstable when $\Re[s(Q_2)]>0$. This condition is similar to (\ref{eq:sq1}) above, but is caused by instability in the warp diffusion coefficient ($Q_2$) and its effect on the warp amplitude. This is distinct from the usual viscous instability as it acts on the warp amplitude rather than the disc surface density. This mode becomes unstable at some critical warp amplitude interval. Fig. \ref{fig:simplified} (right panel) shows the dimensionless growth rates found for this mode. We see that for small $\alpha$ the growth rates of this instability are significantly higher than those found in the first simplified case where we included only the $Q_1$ term. However, for large $\alpha$ there is no instability from $s(Q_2)$. We note that there is a second root in this case which is equal to $-Q_2$, but this root is never unstable within the $\alpha$ formalism. We should also note that in these simplified cases $s$ is real. Thus, the dimensionless growth rates are equal to $s$. Below we will encounter complex values of $s$ in which case the dimensionless growth rate is given by $\Re(s)$.

If this analysis is repeated for (\ref{eq:dr_alt}) as opposed to (\ref{eq:dr}), then (\ref{eq:sq1}) is modified to $s = apQ_1 + a\left|\psi\right|Q'_1$ and thus the \cite{Lightman:1974aa} criterion for thermal-viscous instability is modified from $apQ_1 > 0$ to $apQ_1 > -a\left|\psi\right|Q'_1$. In contrast (\ref{eq:sq2}) remains unchanged.

\begin{figure}
  \begin{center}
   \begin{tabular}{l}
      {\includegraphics[angle=270,scale=.38]{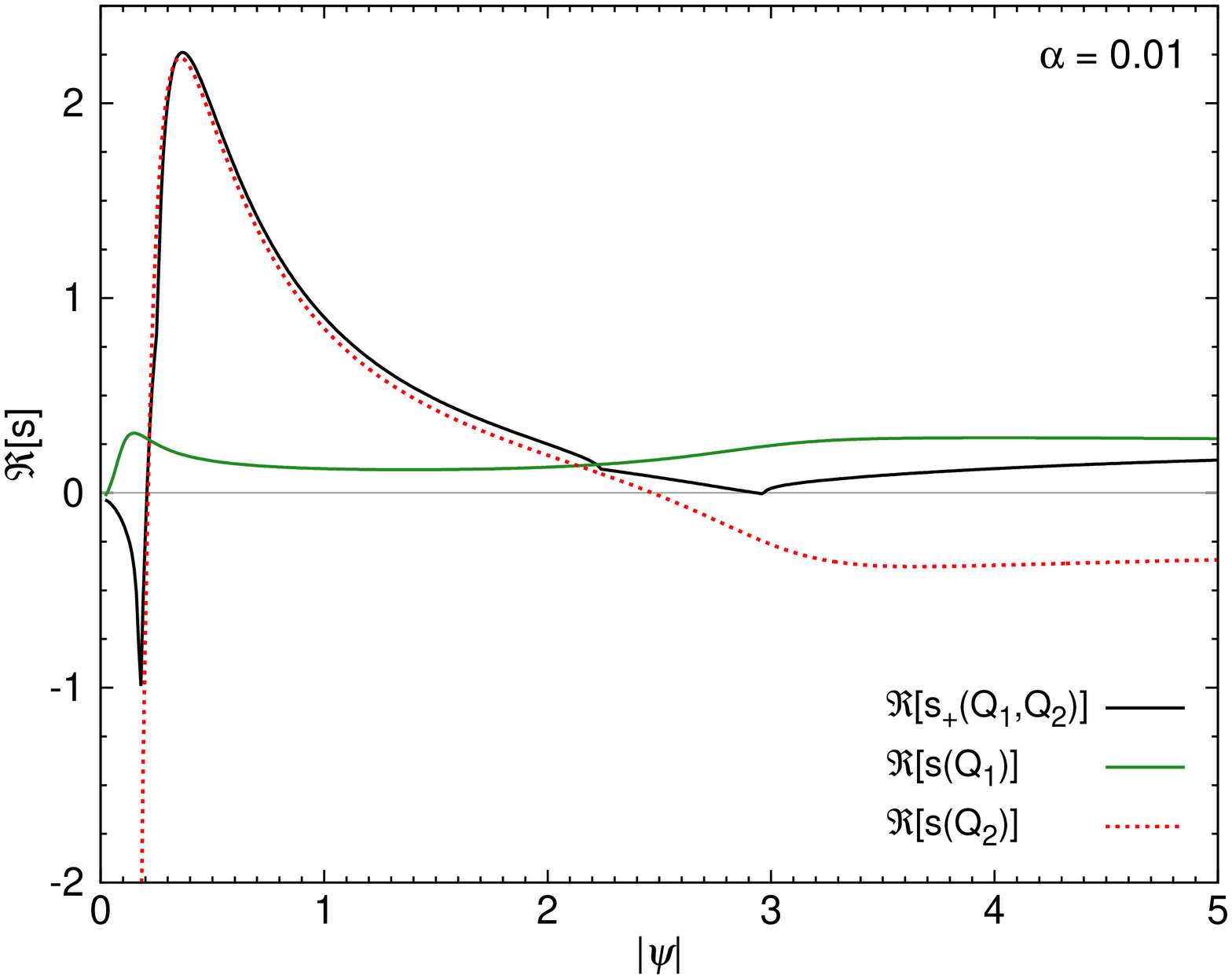}}
      {\includegraphics[angle=270,scale=.38]{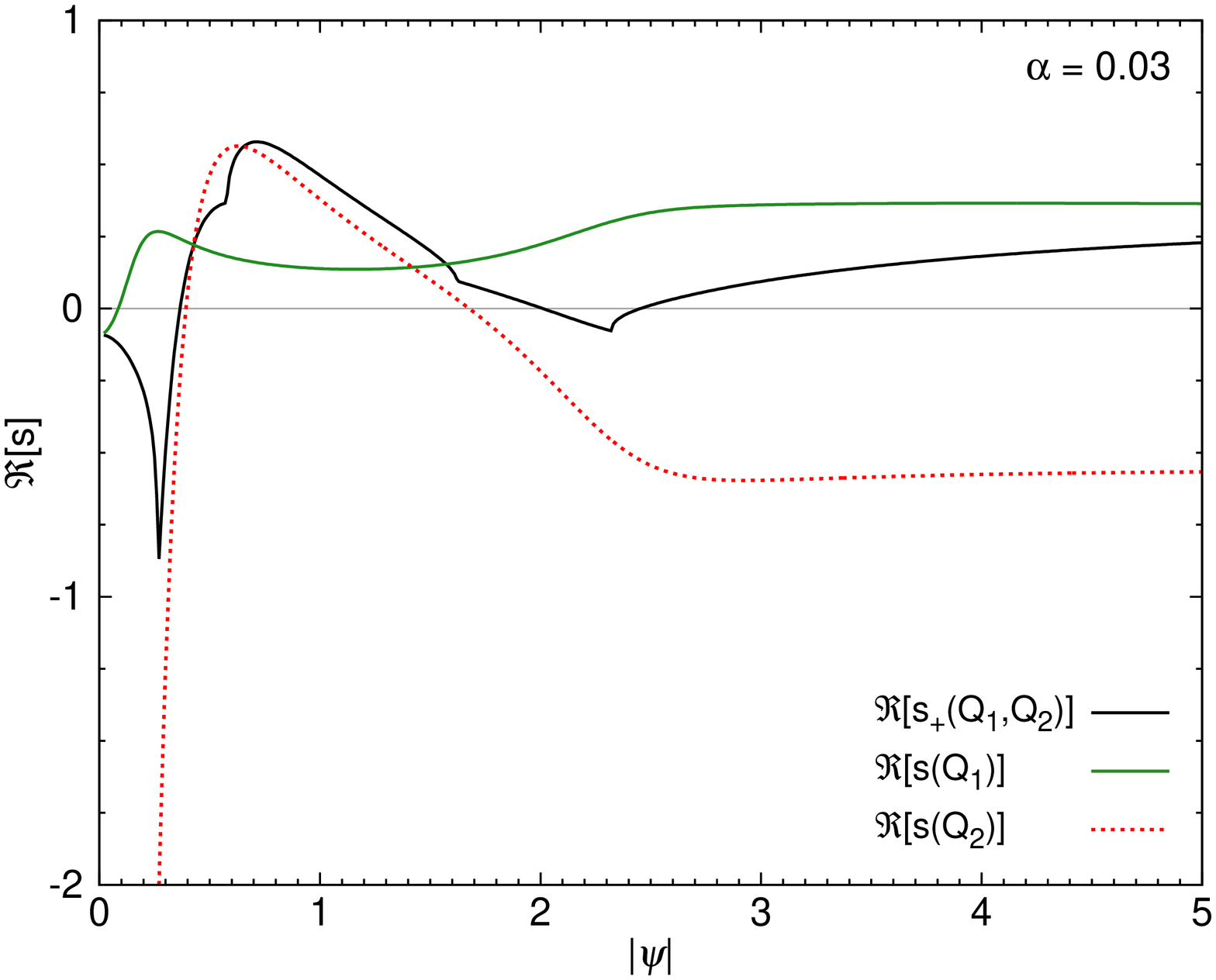}}
         \end{tabular}
   \caption{Here we combine the two panels of Fig.~\ref{fig:simplified} to illustrate the combined effects of the two terms $Q_1$ and $Q_2$. The curves show the dimensionless growth rates $\Re[s_+(Q_1,Q_2)]$ (black solid line) with $|\psi|$ (equation \ref{eq:sq1q2}), compared against $\Re[s(Q_1)]$ (green solid line) and $\Re[s(Q_2)]$ (red dashed line). The grey line represents zero growth rate. For small $\alpha=0.01$ (left panel) we see that instability occurs for all warp amplitudes $\left|\psi\right| > 0.2$, except for a very narrow region around $\left|\psi\right| \approx 2.96$. The growth rate is generally large, except around $\left|\psi\right| \approx 3.0$. The largest growth rates occur between $0.2 < \left|\psi\right| \lesssim 2.5$ where the $Q_2$ term dominates. For larger $\left|\psi\right|$ the $Q_1$ term dominates and the growth rates are sub-dynamical. For intermediate $\alpha=0.03$ (right panel) the same applies except that the the growth rates becomes smaller, and there is a wider band of stability around $\left|\psi\right| = 2.2$. For large $\left|\psi\right|$ the growth rates in the two cases ($\alpha=0.01$ and $\alpha=0.03$) are similar.}
    \label{fig:q1q2}
  \end{center}
\end{figure}
Now we include both $Q_1$ and $Q_2$ terms (i.e. $Q_3=0$) to find a more general criterion for instability. In this case, (\ref{eq:dr}) gives three roots, one of which corresponds to $-Q_2$ again, and is thus stable for any $\left|\psi\right|$. The remaining roots of the cubic correspond to
\begin{equation}
\label{eq:sq1q2}
s_{\pm}(Q_1,Q_2)=\frac{1}{2}\Bigg[a\Big(Q_1 + Q'_1|\psi|\Big)-\Big(Q_2 +Q'_2 |\psi|\Big)\pm\sqrt{\left[a \Big(Q_1 + Q'_1 |\psi|\Big)-\Big(Q_2 + Q'_2 |\psi|\Big) \right]^2 +4a\left[\left(Q_1Q_2 + (Q_1Q'_2 -Q'_1 Q_2)|\psi|\right)\right]}\Bigg],
\end{equation}
 We see that these modes include both instabilities found in the simplified cases above. Inspection of (\ref{eq:sq1q2}) reveals that $\Re(s_+)\geq \Re(s_-)$. Therefore, when the disc is unstable, $s_+$ represents the mode which shows more rapid growth. Fig. \ref{fig:q1q2} shows the variation of $\Re[s_+(Q_1,Q_2)]$ with $|\psi|$.  The disc becomes unstable when  $\Re[s_+(Q_1,Q_2)]>0$. We plot $\Re[s(Q_1)]$ and $\Re[s(Q_2)]$ in the same graph for comparison.  $\Re[s_+(Q_1,Q_2)]$ shows similar behaviour to $\Re[s(Q_2)]$ given by (\ref{eq:sq2}) for $\left|\psi\right| \lesssim 2.5$, and shows similar behaviour to $\Re[s(Q_1)]$ given by (\ref{eq:sq1}) for $\left|\psi\right| \gtrsim 2.5$. The disc becomes unstable at some critical warp amplitude, $|\psi|_c$. For some values of $\alpha$ the disc becomes stable again at larger warp amplitude before becoming unstable again at a second critical warp amplitude. The critical warp amplitudes and the maximum values of the dimensionless growth rates are almost equivalent for $s_+(Q_1,Q_2)$ and  $s(Q_2)$. The $Q_1$ term tends to become dominant only at high warp amplitudes. We should note that $\Re[s_-(Q_1,Q_2)]$ also becomes positive at some critical warp amplitude, but is very close to zero.  As a result, both modes, i.e. $s_+(Q_1,Q_2)$ and $s_-(Q_1,Q_2)$, are unstable after the critical warp amplitude, but $s_+(Q_1,Q_2)$ is always the most unstable.

Finally, if we include all the $Q_i$ terms in the dispersion relation we obtain the full criterion for instability as $\Re\left[s_1(Q_1,Q_2,Q_3)\right] > 0$ with:
\begin{equation}
s_{1}(Q_1,Q_2,Q_3)= \frac{1}{6}\Big[2\mathcal{C}_1+2^{2/3}\Big(\mathcal{C}_2+\mathcal{C}_3^{1/2}\Big)^{1/3}+ 2^{4/3}\Big(\mathcal{C}_2+\mathcal{C}_3^{1/2}\Big)^{-1/3}\mathcal{C}_4\Big]
\label{eq:sfull}
\end{equation}
where
\begin{equation}
\begin{split}
\mathcal{C}_1 =&~~ a \big(Q_1+Q_1' |\psi| \big)-2 \big(Q_2+Q_2' |\psi|\big)
 \end{split}
\end{equation}
\begin{equation}
\begin{split}
 \mathcal{C}_2 = &~~ 3 a \bigg[Q_1' |\psi| \bigg(2 Q_2'^2 |\psi|^2 + 5 Q_2 Q_2' |\psi| - 4 Q_2^2 - 3 Q_3 \Big(Q_3' |\psi| + 4 Q_3\Big)\bigg) + Q_1 \bigg(-Q_2'^2 |\psi|^2 + 2 Q_2 Q_2' |\psi| \\
    &+ 2 \Big(Q_2^2 + 3 Q_3 \big(Q_3' |\psi| + Q_3 \big)\Big)\bigg)\bigg]+2 a^3 \big(Q_1 +Q_1' |\psi|\big)^3 + 3 a^2 \big(Q1^2 -2 Q_1'^2 |\psi|^2 - Q_1 Q_1' |\psi| \big) \big(2 Q_2 + Q_2' |\psi| \big) \\
    &- \big(2 Q_2+Q_2' |\psi|\big) \Big(2 Q_2'^2 |\psi|^2 - Q_2 Q_2' |\psi| - Q_2^2 - 9 Q_3 \big(Q_3' |\psi| + Q_3 \big)\Big) \\
 \end{split}
\end{equation}
\begin{equation}
\begin{split}
\mathcal{C}_3=&~~4 \bigg(3 \Big(Q_2^2+Q_2' Q_2|\psi|-a Q_1 (2 Q_2+Q_2' |\psi| )+Q_3 (Q_3+Q_3' |\psi| )\Big)-\big(2 Q_2+Q_2' |\psi| -a (Q_1+Q_1' |\psi| )\big)^2\bigg)^3\\
&+\Big[2 a^3 (Q_1+Q_1' |\psi| )^3+(2 Q_2+Q_2' |\psi| ) \left(Q_2^2+Q_2'  Q_2|\psi|+9 Q_3^2-2 Q_2'^2 |\psi| ^2+9 Q_3' Q_3 |\psi| \right)\\
&+3 a \bigg(Q_1 \left(2 Q_2^2+2 Q_2' Q_2|\psi|+6 Q_3^2-Q_2'^2 |\psi| ^2+6 Q_3' Q_3 |\psi| \right)+Q_1' |\psi|  \big(-4 Q_2^2+5 Q_2' Q_2|\psi| +2 Q_2'^2 |\psi| ^2\\
&-12 Q_3^2 -3 Q_3' Q_3 |\psi| \big)\bigg)+3 a^2 (2 Q_2+Q_2' |\psi| ) \left(Q_1^2-Q_1' Q_1|\psi|-2 Q_1'^2 |\psi| ^2\right)\Big]^2
 \end{split}
\end{equation}
\begin{equation}
\begin{split}
\mathcal{C}_4= &~~a^2 \big(Q_1+Q_1' |\psi| \big)^2+ Q_2^2+ Q_2' Q_2|\psi| +Q_2'^2 |\psi| ^2 +a \big(2 Q_2+Q_2' |\psi| \big) \big(Q_1-2 Q_1' |\psi| \big)-3 \big(Q_3^2 + Q_3' Q_3 |\psi|\big)
 \end{split}
\end{equation}
\\
The other two roots of the dispersion relation are given by
\begin{equation}
s_{2}(Q_1,Q_2,Q_3)= \frac{1}{6}\Big[2\mathcal{C}_1-2^{-1/3}\big(1-\mathrm{i}\sqrt{3}\big)\Big(\mathcal{C}_2+\mathcal{C}_3^{1/2}\Big)^{1/3}-2^{1/3}\big(1+\mathrm{i}\sqrt{3}\big)\Big(\mathcal{C}_2+\mathcal{C}_3^{1/2}\Big)^{-1/3}\mathcal{C}_4\Big]
\end{equation}
and
\begin{equation}
s_{3}(Q_1,Q_2,Q_3)= \frac{1}{6}\Big[2\mathcal{C}_1-2^{-1/3}\big(1+\mathrm{i}\sqrt{3}\big)\Big(\mathcal{C}_2+\mathcal{C}_3^{1/2}\Big)^{1/3}-2^{1/3}\big(1-\mathrm{i}\sqrt{3}\big)\Big(\mathcal{C}_2+\mathcal{C}_3^{1/2}\Big)^{-1/3}\mathcal{C}_4\Big]\,.
\end{equation}

\begin{figure}
  \begin{center}
      {\includegraphics[angle=270,scale=.4]{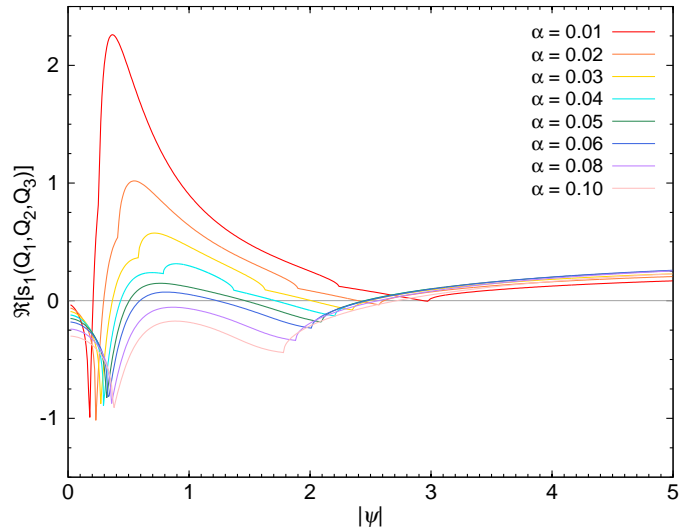}}
      \caption{Here we show the dimensionless growth rates $\Re[s]$ as functions of $\left|\psi\right|$ taking account of all three terms $Q_1$, $Q_2$ and $Q_3$, for different values of $\alpha$. The effect of the $Q_3$ term is negligible for the parameters we have considered (see Figs.~\ref{fig:comparison1} \& \ref{fig:comparison2}).}
      \label{fig:sfull}
  \end{center}
\end{figure}

\begin{figure}
  \begin{center}
      {\includegraphics[angle=270,scale=.4]{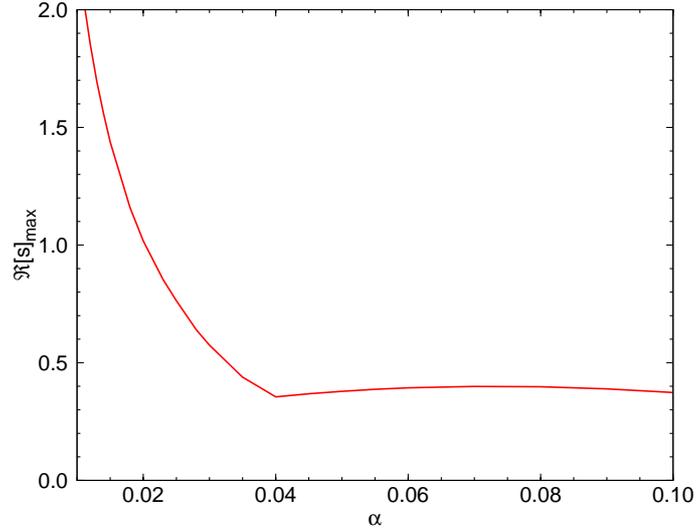}}
      \caption{This plot shows the maximum growth rate of the instability as a function of viscosity parameter $\alpha$. Note that for different values of $\alpha$ the maximum growth occurs for different values of the dimensionless warp amplitude $\left|\psi\right|$. This plot shows that instability occurs for all values of $\alpha$ in this range. For small values of $\alpha$ the maximum growth rate is dynamical. For larger values of $\alpha$ the maximum growth rate is smaller, but still comparable to dynamical ($\sim 0.2\Omega$).}
\label{fig:smax}
  \end{center}
\end{figure}

\begin{figure}
  \begin{center}
      {\includegraphics[angle=270,scale=0.4]{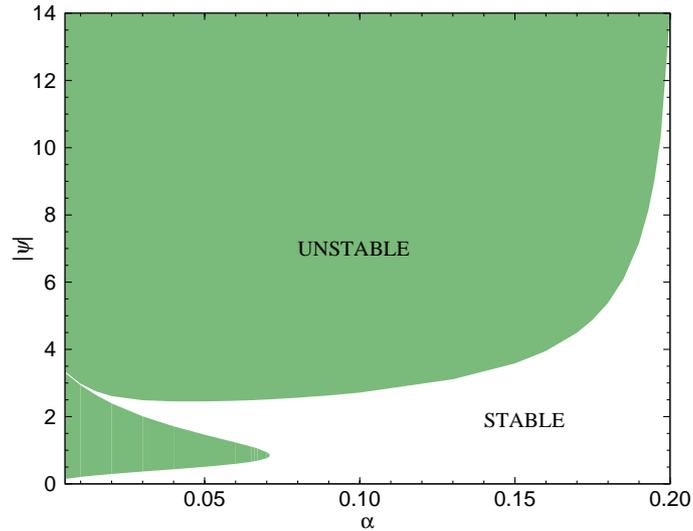}}
    \caption{Stable (white) and unstable (green) regions in the ($\alpha$, $|\psi|$) parameter space. This plot represents the critical warp amplitudes for instability to occur in discs with various $\alpha$ values. We note that a flat disc is always stable (assuming an isothermal equation of state so that there is no Lightman-Eardley-like instability). For $0.01 \le \alpha \le 0.2$ a nearly flat disc with $\left|\psi\right| \lesssim 0.1$ is also always stable. For a given value of $\alpha$ there is always a minimum value of the warp amplitude which gives rise to instability.}
\label{fig:psi_c}
  \end{center}
\end{figure}

These roots are either all real, or there can be one real root and two complex conjugates. For determining instability we are interested in the real roots and the real parts of the complex roots. While there is always one real root, the other two roots can swap between real or complex-conjugates for different values of $\left|\psi\right|$. One root is always stable. Of the other two roots one always gives the fastest growth (i.e. when the real part is positive it shows the largest value) and this root is given by (\ref{eq:sfull}). For some values of $\left|\psi\right|$ this root is a complex conjugate and thus the other root has the same growth rate. In Fig. \ref{fig:sfull}, we show the dimensionless growth rates of the instability found from the full criterion (\ref{eq:sfull}) for different values of $\alpha$. We see that the disc becomes unstable for sufficiently large $|\psi|$ values. The critical warp amplitudes where the disc becomes unstable are smaller and the growth rates of the instability are higher for low $\alpha$.
In Fig. \ref{fig:smax}, we plot the maximum values of the dimensionless growth rates of instability ($\Re[s]_{\rm max}$) against $\alpha$. We see that $\Re[s]_{\rm max}$ values are inversely proportional to $\alpha$. Finally, the critical warp amplitudes for instability are plotted against $\alpha$ in Fig. \ref{fig:psi_c}. This figure shows the stable and unstable regions in the ($\alpha$, $|\psi|$) parameter space.

\begin{figure}
  \begin{center}
      {\includegraphics[angle=270,scale=0.38]{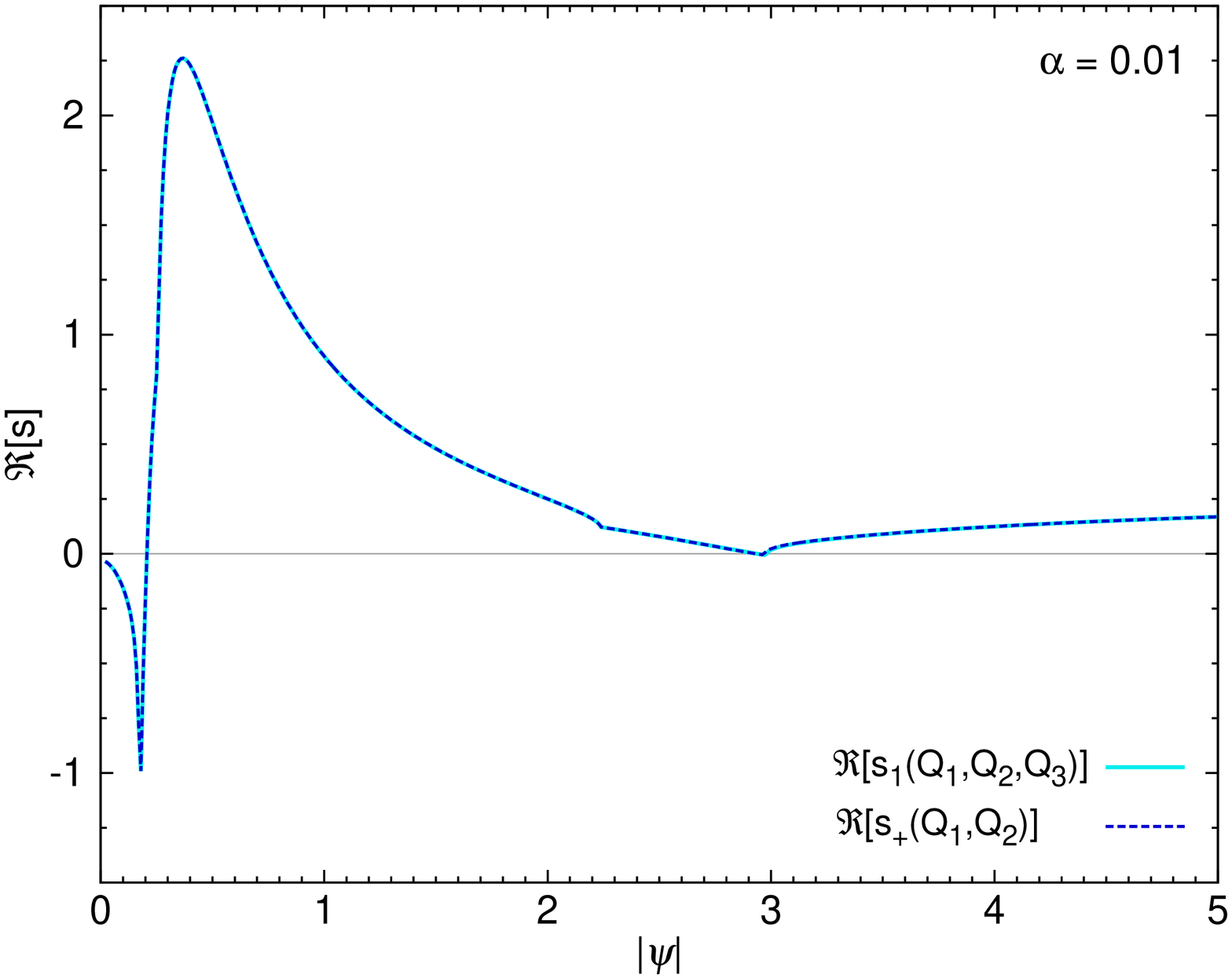}}
      {\includegraphics[angle=270,scale=0.38]{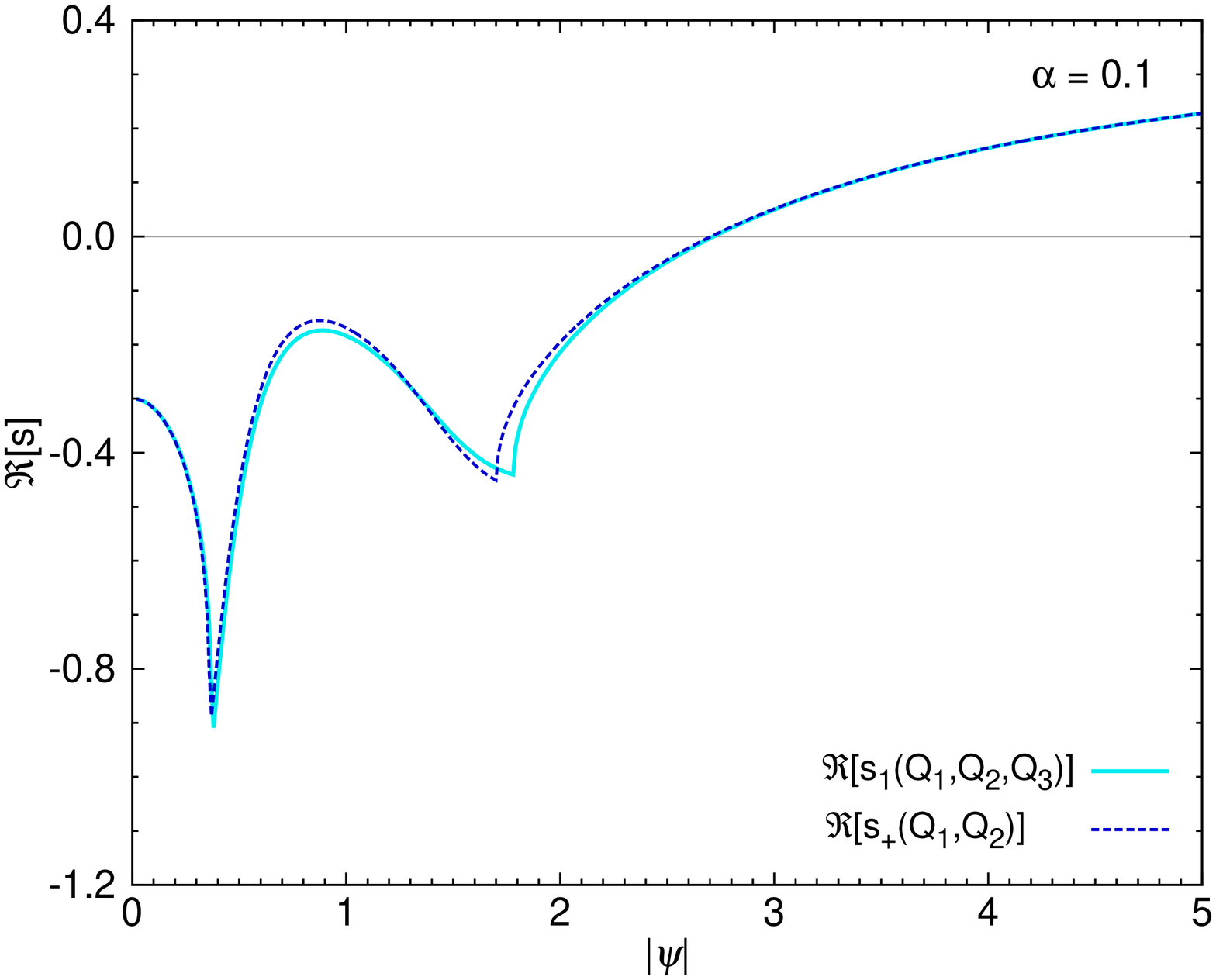}}
    \caption{Comparison of the full (\ref{eq:sfull}) and simplified criteria (\ref{eq:sq1q2}) for instability for $\alpha = 0.01$ and $0.1$. The grey line represents zero growth rate. These plots show that the simplified criteria are an extremely good approximation to both the location of the critical warp amplitude(s) and the dimensionless growth rate as a function of warp amplitude.}
\label{fig:comparison1}
  \end{center}
\end{figure}

\begin{figure}
  \begin{center}
      {\includegraphics[angle=270,scale=0.4]{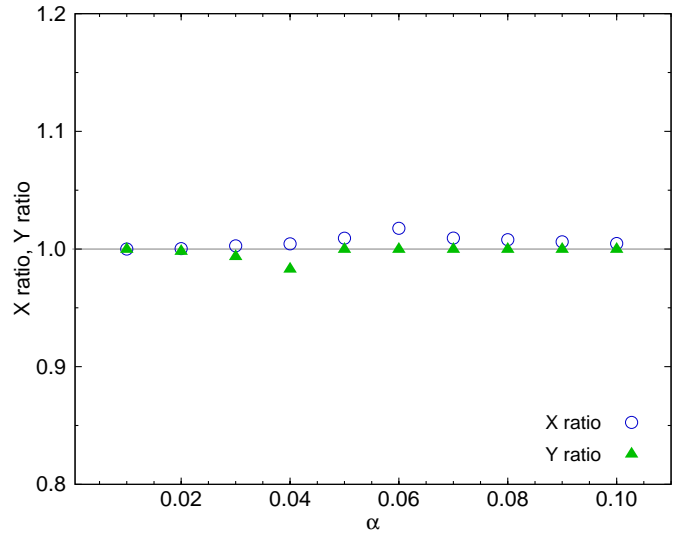}}
           \caption{$X=|\psi|_c\Re[s(Q_1,Q_2,Q_3)]/|\psi|_c\Re[s(Q_1,Q_2)]$ and $Y=\Re[s_{\rm max}(Q_1,Q_2,Q_3)]/\Re[s_{\rm max}(Q_1,Q_2)]$ ratios for different $\alpha$ values. The grey line represents unity. Thus, the critical warp amplitudes and the maximum values of the dimensionless growth rates are captured by the simplified root (\ref{eq:sq1q2}) to high accuracy for all of the $\alpha$ values we have explored.}
\label{fig:comparison2}
  \end{center}
\end{figure}

If we compare (\ref{eq:sq1q2}) and (\ref{eq:sfull}), we see that the inclusion of $Q_3$ produces a big difference in the formula, but does not have a strong impact on the numerical results. In Fig. \ref{fig:comparison1}, we compare two criteria, (\ref{eq:sq1q2}) where $Q_3 = 0 $ and (\ref{eq:sfull}) where $Q_3 \neq 0$. We see that the inclusion of $Q_3$ has a very small effect on the critical warp amplitudes and the growth rates. We also plot the ratios of the $|\psi|_c$ and $\Re[s]_{\rm max}$ found for the two different cases, $s(Q_1,Q_2)$ \& $s(Q_1,Q_2,Q_3)$. Fig. \ref{fig:comparison2} shows the values of $X$ and $Y$ as a function of $\alpha$, where $X=|\psi|_c\Re[s(Q_1,Q_2,Q_3)]/|\psi|_c\Re[s(Q_1,Q_2)]$ and $Y=\Re[s_{\rm max}(Q_1,Q_2,Q_3)]/\Re[s_{\rm max}(Q_1,Q_2)]$. The solutions are in excellent agreement. Therefore, (\ref{eq:sq1q2}) can be used to decide whether the disc is stable or unstable, at least for the parameters we have explored.

In conclusion, the criteria for instability can be expressed from (\ref{eq:sq1q2}) as below:

\begin{equation}
\label{eq:crit1}
\rm If \,\,\,\,\,\left[a\cfrac{\partial}{\partial \left|\psi\right|}\left(Q_1\left|\psi\right|\right)-\cfrac{\partial}{\partial \left|\psi\right|}\left(Q_2\left|\psi\right|\right)\right]>0,\,\rm the \,\rm disc \,\rm is \,\rm unstable,
\end{equation}

\begin{equation}
\label{eq:crit2}
\rm or \,\rm if \,\,\,\,\ \left[a\cfrac{\partial}{\partial \left|\psi\right|}\left(Q_1\left|\psi\right|\right)-\cfrac{\partial}{\partial \left|\psi\right|}\left(Q_2\left|\psi\right|\right)\right]<0,\,\,\,\,\,\,\,\,\,\,\rm and \,\,\,\,\,\,\,\,\,\,
4a\left[\left(Q_1Q_2 + (Q_1Q'_2 -Q'_1 Q_2)|\psi|\right)\right]>0,\,\rm the \,\rm disc \,\rm \,is \,\rm also\, \rm unstable.
\end{equation}


\section{Heuristic picture of the viscous-warp instability}
\label{sec:heuristic}
To elucidate the fundamental physics of this viscous-warp disc instability from the complex analysis described above, we present here a simplified analysis. Starting from (\ref{eq:here}) above, we simplify the picture aggressively. In a warped disc with $\alpha \ll 1$, the smoothing of the disc warp occurs faster than mass is transported radially \citep{Papaloizou:1983aa}. So we assume that the important torque is the diffusive torque relating to $\nu_2$. Our exploration of the analysis above confirms this is a sensible approach in many, but not all, cases. Thus
\begin{equation}
  \label{eq:simp1}
  \frac{\partial}{\partial t}\left(\Sigma r^2\Omega\mathbfit{l}\right) \sim \frac{1}{r}\frac{\partial}{\partial r}\left(Q_2\Sigma c_{\rm s}^2r^3\frac{\partial \mathbfit{l}}{\partial r}\right)\,.
\end{equation}
Interpreting this as a diffusion equation implies that the flux of misaligned angular momentum is
\begin{equation}
  \mathbfit{F}_{\rm mis} = Q_2\Sigma c_{\rm s}^2r^3\frac{\partial \mathbfit{l}}{\partial r}
\end{equation}
with rate of transfer $\sim \left|\mathbfit{F}_{\rm mis}/\Sigma r^2 \Omega\right| = (1/2)\nu_2\left|\psi\right|$. Connecting this with classical instability of a diffusion equation yields the criterion for instability as
\begin{equation}
  \label{eq:simpcrit}
  \frac{\partial}{\partial \left|\psi\right|}\left(\nu_2\left|\psi\right|\right) < 0\,.
\end{equation}
This can also be trivially recovered by performing a stability analysis on (\ref{eq:simp1}), with $\mathbfit{l} \rightarrow \mathbfit{l}+\delta \mathbfit{l}$, $\delta Q_2 = (\partial Q_2/\partial \left|\psi\right|)\delta\left|\psi\right|$ and assuming $\delta\Sigma = 0$.

The criterion (\ref{eq:simpcrit}) can be interpreted as stating if $\nu_2\left(\left|\psi\right|\right)$ is such that the maximum diffusion rate is not located at maxima in warp amplitude, then local maxima in warp amplitude will grow, and the disc will break. As this will result in rapid transfer of mass out of this region due to the large warp amplitude implying large torques, this will also be realised by a significant drop in local surface density. This resembles the Lightman-Eardley viscous instability but for a warped disc with the warp amplitude playing the role of the surface density. Such behaviour is impossible if the effective viscosities are constant with varying warp amplitude. This is confirmed in our analysis above which yields no unstable solutions if $Q_i$ are independent of $|\psi|$. This is consistent with the results of \cite{Nixon:2012aa}\footnote{Note that in \cite{Nixon:2012aa} when using the $Q_i$ coefficients from \cite{Ogilvie:1999aa} we smoothed the viscosity coefficients over the neighbouring few rings to avoid what was perceived there as numerical issues. Actually this behaviour was the physical instability discussed here.}. Finally we note that when the correct limits are applied, the full stability analysis presented above reproduces the {\it ad hoc} answer derived here (see equation~\ref{eq:sq2} above).

\section{Application to disc tearing}
\label{sec:tearing}
In this work we have focused on disc `breaking', where an isolated already-warped disc (i.e. with no external torque) evolves to create a discontinuity between two or more planes, with a series of low-density orbits connecting them over a small radial extent. We have revealed above that this occurs due to a viscous-warp instability, which occurs when either (\ref{eq:crit1}) or (\ref{eq:crit2}) are satisfied.
In many physical systems, the disc is subject to an external torque often occurring due to precession (e.g. the Lense-Thirring effect). In this case, the warp initially grows with time. If the disc `breaks' under these conditions creating disconnected rings which keep precessing effectively independently, then this is what we have called disc `tearing' in previous papers \citep[e.g.][]{Nixon:2012ad,Nealon:2015aa,Dogan:2015aa}.

It is of great interest to connect the analysis in this work to the idea of disc tearing. Unfortunately at first appearances, we cannot perform the same analysis as above for disc tearing, where an external forced precession is imposed, as the background unperturbed state is no longer necessarily an equilibrium solution on short timescales. It may be that this is unimportant, and an approximate solution can be found that is accurate enough. We will check this in future work. For our purposes here, we repeat that the analysis above is a local analysis which takes a given warped disc shape and determines its stability. For disc tearing we need to consider how the disc gets into a warped configuration, and at what stage during that process instability occurs.

Here we can make a reasoned guess at what the disc tearing criterion is likely to be. Above we have derived, for various system parameters, the critical warp amplitude $\left|\psi\right|_c$ at which the disc becomes unstable. It would seem reasonable that a forced disc would tear up if there is no $\left|\psi\right| < \left|\psi\right|_c$ at which the torque attempting to flatten the disc ($\propto Q_2\left|\psi\right|$) is stronger than the torque induced by e.g. precession. As discussed in \cite{Dogan:2015aa}, without the knowledge of the value of $\left|\psi\right|_c$ this calculation cannot be done {\it a priori} without performing a full numerical simulation. However, our analysis enables exactly that. We will test this in the future by comparing this reasoning with numerical simulations.

We note that given the viscous torque and precession torque (in the Lense-Thirring case) have no dependence on inclination angle, this criterion at first appears at odds with previous simulations which show e.g. a strong inclination angle dependence. There is also no dependence on the disc thickness, but previous simulations show that thinner discs are easier to tear. The answer to this is likely to be that the warp amplitude cannot be larger than a value dependent on the inclination angle and $H/R$. This can be see by the following argument. First the warp amplitude is
\begin{equation}
  \left|\psi\right| = r\left|\frac{\partial \mathbfit{l}}{\partial r}\right|\,,
\end{equation}
and the disc unit tilt vector is
\begin{equation}
  \mathbfit{l} = \left(\cos\gamma\sin\beta,\sin\gamma\sin\beta,\cos\beta\right)\,,
\end{equation}
where $\gamma$ is the disc twist and $\beta$ is the disc tilt \citep[e.g.][]{Nixon:2012ac}. Let's assume $\gamma = 0$, and a maximal warp from $\beta$ to zero (the same result, equation \ref{eq:psimax} below, is arrived at if $\beta$ is assumed a constant and $\gamma$ varies from $0$ to $\pi$). Thus
\begin{equation}
  \delta\mathbfit{l} = \left(\sin\beta,0,\cos\beta-1\right)
\end{equation}
 which has magnitude $(2-2\cos\beta)^{1/2}$.
We can also assume that $\delta r$ cannot become smaller than $H$, and thus the maximum warp amplitude is given by
\begin{equation}
  \label{eq:psimax}
  \left|\psi\right|_{\rm max} \sim \left(2-2\cos\beta\right)^{1/2}\frac{R}{H} \sim \beta \frac{R}{H}\,.
\end{equation}
Therefore discs which are already close to alignment, or are thick, are difficult to tear, as the maximum warp amplitude is small and potentially less than the critical warp amplitude required for instability.

\section{Conclusions}
\label{sec:conclusions}
We have derived the criterion for isolated warped discs to break. This occurs physically due to viscous anti-diffusion of the warp amplitude. The resulting large torques that are induced lead to the disc surface density lowering in the unstable regions. Thus the instability leads to a succession of orbits with planes that change rapidly with radius, with only a small amount of mass present. This is a local instability so it is not likely to directly effect global disc properties such as the central accretion rate. However, this instability underlies the process of disc tearing which has the capacity to dramatically alter the instantaneous accretion rate and the observable properties of the disc on short timescales.

The instability is governed by a complicated criterion (\ref{eq:sfull}), but we have shown that the criterion can be simplified in most cases to a more manageable criteria given by (\ref{eq:crit1}) and (\ref{eq:crit2}). This simply implies that the diffusion of the warp must be maximal at the location of the maximum warp amplitude otherwise the warp will grow until the disc disconnects on scales of order the disc thickness $\sim H$. In future works we will perform this analysis for non-Keplerian discs, connect our results with numerical simulations and include a more sophisticated thermal treatment.

For discs with moderate values of $\alpha \sim 0.01-0.1$ and reasonably thin $H/R \lesssim \alpha$, our work supports the idea of discs being viscously unstable to disconnecting into discrete planes. This process underpins the state-transition cycle suggested by \cite{Nixon:2014aa}. The natural link between the viscous instability in planar discs, which underlies the thermal-viscous outburst cycle observed in many types of binary systems (many of which are thought to contain warped discs), and the viscous instability of warped discs is appealing. This may be a way to introduce hysteresis through a change in the disc stability with warping. We will explore this in future work.

\section*{Acknowledgments}
We thank the referee for providing a helpful and detailed report. SD gratefully acknowledges the warm hospitality of the Theoretical Astrophysics Group at University of Leicester during her visit. SD is supported by the Turkish Scientific and Technical Research Council (T\"{U}B\.{I}TAK - 117F280). CJN is supported by the Science and Technology Facilities Council (grant number ST/M005917/1). The Theoretical Astrophysics Group at the University of Leicester is supported by an STFC Consolidated Grant.

\bibliographystyle{mnras}
\bibliography{nixon}

\bsp
\label{lastpage}
\end{document}